\documentclass[RNAAS]{aastex62}

\renewcommand{\thefootnote}{\fnsymbol{footnote}}

\newcommand{\gsim}{\lower.7ex\hbox{\ensuremath{\;\stackrel{\textstyle>}{\sim}\;}}}
\newcommand{\lsim}{\lower.7ex\hbox{\ensuremath{\;\stackrel{\textstyle<}{\sim}\;}}}

\begin{document}

\title{A TESS Search for Distant Solar System Planets: A Feasibility Study}

\correspondingauthor{Matthew~J.~Holman}
\email{mholman@cfa.harvard.edu}
\author[0000-0002-1139-4880]{Matthew~J.~Holman}
\affiliation{Harvard-Smithsonian Center for Astrophysics, 60 Garden St., MS 51, Cambridge, MA 02138, USA}

\author[0000-0001-5133-6303]{Matthew~J.~Payne} 
\affiliation{Harvard-Smithsonian Center for Astrophysics, 60 Garden St., MS 51, Cambridge, MA 02138, USA}

\author[0000-0001-5449-2467]{Andr{\'a}s~P{\'a}l}
\affiliation{Konkoly Observatory, Research Centre for 
Astronomy and Earth Sciences, Konkoly-Thege M. \'ut 15-17, Budapest, H-1121, 
Hungary}
\affiliation{Department of Astronomy, Lor\'and E\"otv\"os University, 
P\'azm\'any P. stny. 1/A, Budapest H-1117, Hungary}


\section{} 



TESS~\citep{2015JATIS...1a4003R} 
monitors the sky through four cameras, each imaging $24\arcdeg\times 24\arcdeg$, with $21\farcs1$ pixels.
During its two-year mission, TESS observes most of the sky, omitting within $\sim\pm6\arcdeg$ of the ecliptic.  Each `sector' is observed for two 13.7~day TESS orbits.
In its approved extended mission, TESS will re-observe most of the sky, including two-thirds of the ecliptic, increasing sky coverage to $\sim94\%$.

The TESS cameras operate in shutter-less mode, taking 
2~s exposures.  
These are combined 
into pre-selected regions recorded at 2-min cadence and full-frame images (FFIs) recorded at 30-min cadence. 
Cosmic ray rejection reduces the effective FFI integration time to 1440~s.

Faint objects can be detected by combining FFIs.  
With TESS, the signal from a solar-color source with magnitude $I_C$ is $S=t_{\rm exp}\,A\,s_0\times10^{-0.4I_C}$, 
where $t_{\rm exp}$ is the exposure time, $A=69\,\mathrm{cm}^2$ is the effective area, 
$s_0=1.45 \times 10^6 \mathrm{ph} \,s^{-1} \mathrm{cm}^{-2}$, and the bandpass is 600-1000~nm
\citep{2015ApJ...809...77S}.  

We estimate the noise as 
$N = \left[S + n_{\rm pix}\,Z_L\,t_{\rm exp} + n_{\rm pix}\,n_{r} \,R_N^2\right]^{1/2}$,
where $n_{\rm pix}$ is the number of aperture pixels, the zodiacal light is $Z_L \sim 47-135\,\mathrm{ph}\,\mathrm{pix}^{-1}\,s^{-1}$, the number of readouts 
is $n_r$,
and the read noise is $R_N \sim 10\,\mathrm{e}^-/\mathrm{pix}$
~\citep{2015ApJ...809...77S}.  
At faint magnitudes, zodiacal light is the dominant noise source.  
The aperture is dictated by the pixel response function (PRF), with 90\% ensquared energy within 4 pixels~\citep{2015JATIS...1a4003R}. 

\begin{figure*}
    \begin{minipage}[b]{\textwidth}
    \centering
    \includegraphics[angle=0,width=0.95\columnwidth]{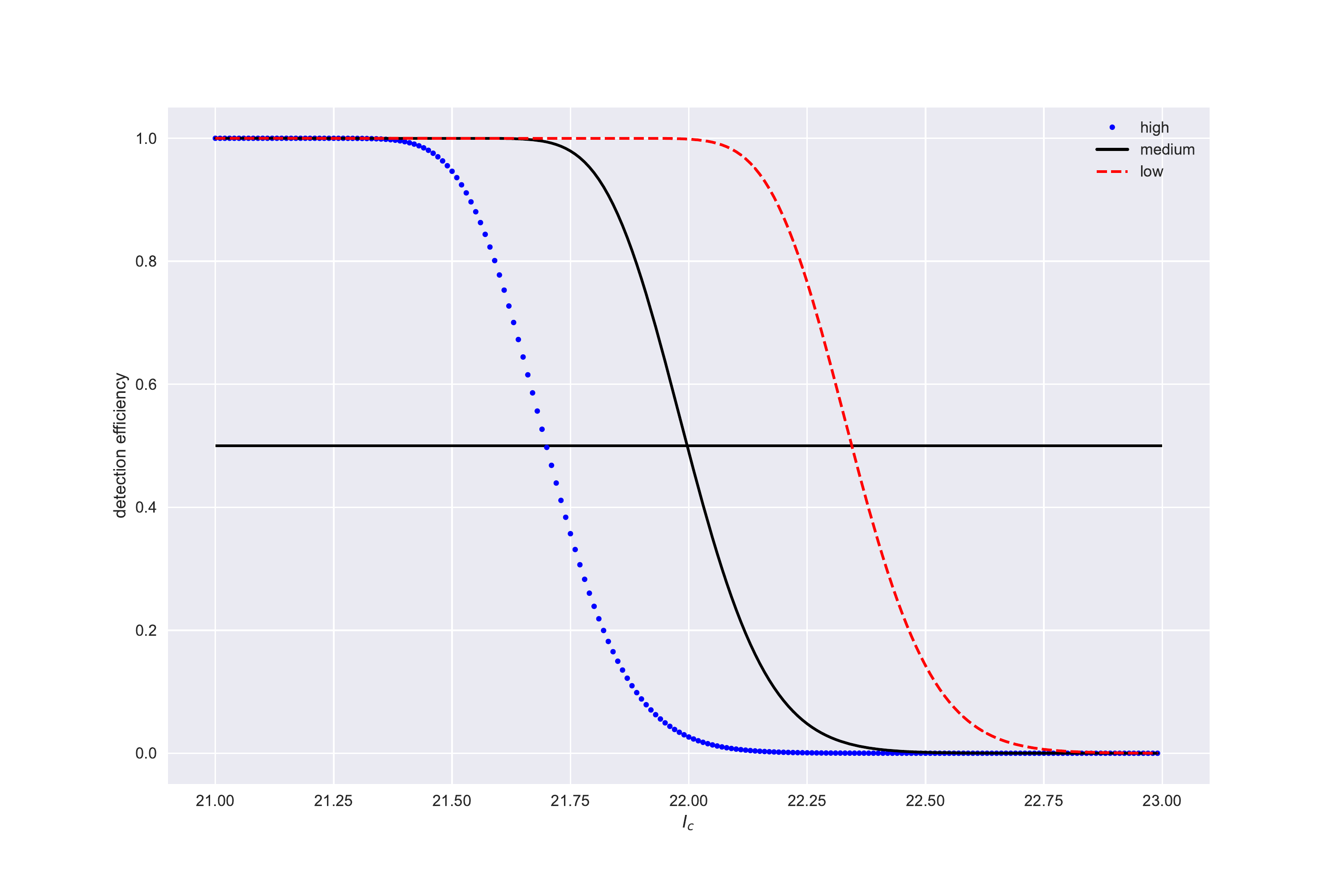}
    \hspace*{-0.0cm}\includegraphics[angle=0,width=0.25\columnwidth]{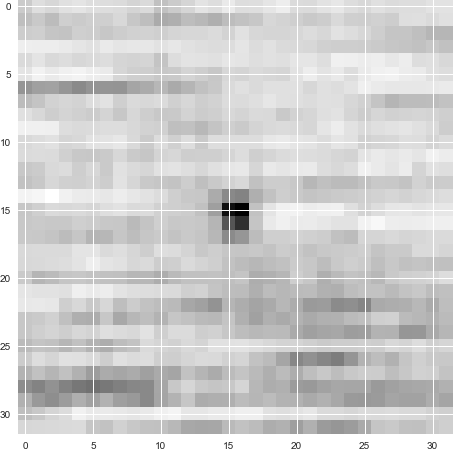}
    \hspace*{1cm}\includegraphics[angle=0,width=0.25\columnwidth]{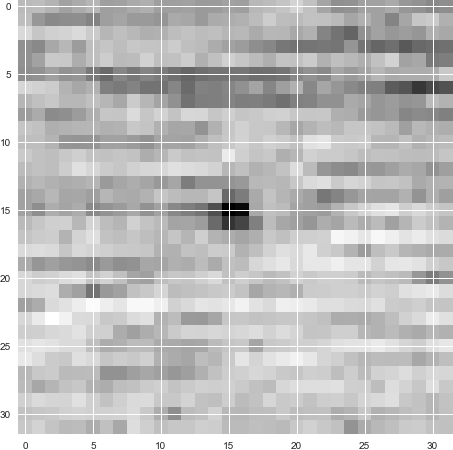}
    \hspace*{1cm}\includegraphics[angle=0,width=0.25\columnwidth]{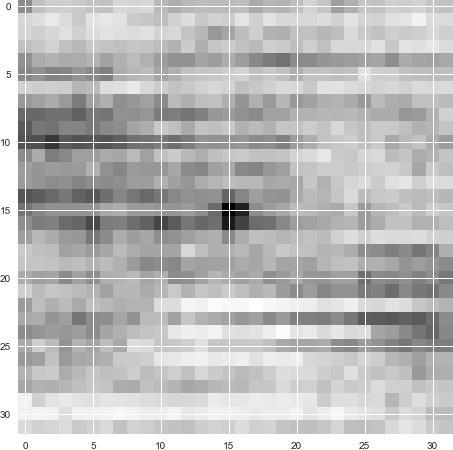}
    \caption{
        {\bf Top:} Predicted detection efficiency $f = \frac{1}{2}\left[1.0 - \mathrm{erf}(X/2)\right]$, where $X=n_\sigma-\mathrm{S/N}$, and $n_\sigma=5$.  The curves correspond to zodiacal light values of $47\,\mathrm{ph}\,\mathrm{pix}^{-1}\,s^{-1}$, $135\,\mathrm{ph}\,\mathrm{pix}^{-1}\,s^{-1}$, and $270\,\mathrm{ph}\,\mathrm{pix}^{-1}\,s^{-1}$.  The third value doubles the maximum estimated zodiacal light value, to account for unmodeled noise.  We assume $n_{pix}=4$.
        {\bf Bottom:} Differential TESS data stacked around the predicted locations of 
        (90377) Sedna ($I_C\sim20.2$),
        2015 BP519 ($I_C\sim21.6$), and
        2015 BM518 ($I_C\sim21.6$), left to right.
        Their $\mathrm{S/N}$ values are $11.1$, $8,7$, and $7.3$, respectively.
        Images created and processed using FITSH ~\citep{Pal.2012}.
}
    \label{f:known}
    \end{minipage}
\end{figure*}

Figure~\ref{f:known} displays the resulting detection efficiency curves.
Combining $\sim\!1,300$ exposures from a TESS sector, gives a $50\%$ detection threshold of $I_C\sim 22.0\pm0.5$.  
TESS will observe portions of the sky for $\gg27$ days, increasing the depth.

\section{Digital Tracking}
\label{s:SaS}
%
The 
curves in Figure~\ref{f:known} also apply to moving objects.
Given a {\it known} orbit, one can predict an object's location in a series of background-subtracted TESS FFIs and sum the flux.
Figure~\ref{f:known} demonstrates this for three TNOs.

To discover new objects, with unknown trajectories, we can try \emph{all possible orbits}!
Previous searches have demonstrated the power of `digital tracking' to detect solar system bodies significantly fainter than single exposure limits~\citep{1998AJ....116.2042G, Gladman.2001, 2004Natur.430..865H,Bernstein04}.  

One can shift a series of images to compensate for the parallax.
The remaining proper motion then yields straight line trajectories.  
By shifting the parallax-compensated images along all plausible linear velocities, one can sum the flux and search for significant peaks in the signal. 
One only needs an \emph{approximate} orbit.  
In the case of TESS, the size of the PRF sets the precision with which the signal in the images must be aligned.

The basis of \citet{Bernstein00} simplifies an exhaustive orbit search.  
\citet{Bernstein04} demonstrated this with an HST survey for extremely faint TNOs.  The key parameters are the parallax constant $\gamma=1/d$ for distance $d$, scaled radial velocity $\dot\gamma=\dot d/d$, and transverse angular velocities $\dot\alpha$ and $\dot\beta$.
For short time spans, $\dot\gamma\approx0$.

%

The sky-plane resolution, $P$, is similar to the pixel scale.
%
The number of angular velocity bins required
is
$$N_{\dot\alpha} = N_{\dot\beta}
= 2\dot\alpha_{\rm max}/\Delta\dot\alpha 
\approx 100 \left(\frac{T}{27\,\mathrm{day}}\right) \left(\frac{d}{25\, \mathrm{au}}\right)^{-1.5} \left(\frac{P}{21 ''}\right)^{-1} , $$ 
\noindent where $\alpha_{\rm max}$ is the maximum bound angular velocity, $T$ is the span of the observations, and the angular velocity resolution is $\Delta\dot\alpha\lsim\,P/T$. 
The number of distance bins is small for TESS~\citep{Bernstein00,Holman18}:
$$N_{\gamma} = 
\gamma / \Delta\gamma 
\sim 
\left(\frac{T}{27\,\mathrm{day}}\right)
\left(\frac{P}{21''}\right)^{-1}
\left(\frac{d}{25\,\mathrm{au}}\right)^{-1}
\sim 1. 
$$
\noindent The total number of operations is
$$N_{\rm op} 
= 
N_{\rm sec} N_{\dot\alpha}  N_{\dot\beta} N_{\gamma} N_{\rm pix} N_{\rm exp} 
\sim 
5\times10^{15}
\left(\frac{N_{\rm sec}}{26\,\mathrm{sectors}}\right)
\left(\frac{T}{27\,\mathrm{day}}\right)^3
\left(\frac{P}{21''}\right)^{-3}
\left(\frac{d}{25\,{\rm au}}\right)^{-4}
\left(\frac{N_{\rm exp}}{1,300}\right)
\left(\frac{N_{\rm pix}}{16\,\mathrm{Mpix}}\right),
$$
where $N_{\rm sec}$ is the number of sectors, $N_{\rm pix}\propto P^{-2}$ is the number of pixels, 
and $N_{\rm exp}$ the number of exposures. 
The \citet{Bernstein04} search required $N_{\rm op}\sim10^{16}$ operations.

TESS can detect objects at $d \lesssim 900\left(\frac{5\,\mathrm{pix}}{n_{p}}\right)\,{\rm au}$, assuming a minimum $n_p\sim5$ pixel displacement.  The hypothesized Planet Nine~\citep{Trujillo.2014,Brown.2016}, has an expected magnitude of $19 < V <24$~\citep{Fortney.2016,2019arXiv190210103B}, raising the possibility that TESS could discover it! 

The expected yield of a TESS search for TNOs and Centaurs will be the subject of a future investigation.


%


%

\section*{Acknowledgments}
We acknowledge 
Charles Alcock,  
Michele Bannister,  
Jason Eastman, 
Adina Feinstein,
Wes Fraser, 
Pedro Lacerda, 
Ben Montet,
George Ricker, 
Scott Sheppard, 
David Trilling,
Chad Trujillo, 
Deborah Woods, Roland Vanderspek, 
and grants from NASA (NNX16AD69G, NNX17AG87G), Hungary (NKFIH-K125015), and 
the Smithsonian.

\bibliographystyle{aas-compact}

\bibliography{references}

\end{document}